Springer series *Resistance to Targeted Anti-Cancer Therapeutics* spearheaded by Dr. Benjamin Bonavida—

*Edited volume: TRAIL, Fas Ligand, TNF and TLR3 in Cancer*

**Chapter: System modeling of receptor-induced apoptosis.**

**François Bertaux[1, 2, 3], Dirk Drasdo[1] and Grégory Batt[3]**

[1] INRIA Paris, 2 rue Simone Iff, 75589 Paris cedex 12, France
[2] Department of Mathematics, Imperial College London, London SW7 2AZ, UK
[3] Inria Saclay – Ile-de-France, 91120 Palaiseau, France

Corresponding authors:

François Bertaux, E-mail: f.bertaux@imperial.ac.uk

Grégory Batt, E-mail: Gregory.Batt@inria.fr

**Table of Contents**

**Abstract**

Receptor-induced apoptosis is a complex signal transduction pathway involving numerous protein/protein interactions and post-transcriptional modifications. The response to death receptor stimulation varies significantly from one cell line to another and even from one cell to another within a given cell line. In this context, it is often difficult to assess whether the molecular mechanisms identified so far are sufficient to explain the rich quantitative observations now available, and to detect possible gaps in our understanding. This is precisely where computational systems biology approaches may contribute. In this chapter, we review studies done in this direction, focusing on those that provided significant insight on the functioning of this complex pathway by tightly integrating experimental and computational approaches.

# 1. Introduction

Apoptosis is a form of programmed cell death conserved among metazoans playing a central role in development and involved in many diseases. Notably, most successful non-surgical cancer therapies eventually result in the activation of apoptosis in cancer cells [1]. Apoptosis can be triggered internally (via an 'intrinsic' pathway) following DNA damage or other intrinsic stimuli, or externally (via an 'extrinsic' pathway) following the binding of 'death' ligands to 'death' receptors. Receptor-induced apoptosis raised therapeutical interest as anti-cancer strategy for at least two reasons. Firstly it can be highly selective for certain cell types, ideally targeting only cancer cells. Secondly it does not require a functional p53 protein, which is frequently mutated in tumor cells, providing resistance to chemotherapeutic drugs relying on the DNA damage response. Several death ligand/receptor pairs exist. While TNF$\alpha$ (and its receptors TNFRs) and CD95L (and its receptor CD95) were discovered first, TRAIL (and its receptors DRs) has the highest selectivity towards cancer cells [2]. From a largest perspective, the latter is also a reference system illustrating how complex circuits involving graded and competing molecular signals can generate binary decisions. Because of its high interest, both for systems biology and therapeutics, tremendous research efforts have been done to better understand its functioning [3–8].

The control and regulation of apoptosis involve many genes whose products mediate numerous protein/protein interactions, post-translational modifications, transcriptional regulations, etc., yielding a highly complex picture. The sensitivity of cells to a given death ligand stimulation is multi-factorial, and the effect of genetic perturbations on cell survival is highly context-dependent. As a result, the interpretation of results obtained on a specific cell line and for a few genetic perturbations or conditions is delicate and cannot be readily generalized. The system model paradigm is well suited to deal with this complexity. Computational approaches attempt to integrate known mechanisms and interactions into mathematical models, whose predictions can be used to propose new experiments for model validation. When applied to apoptosis, system level modeling, which started approximately 15 years ago, was indeed instrumental in improving our understanding of this complex process. This was achieved by tightly integrating data of increasing quality and by increasing the scope and/or level of details of models. Excellent reviews discuss these modeling works [9–15].

Despite these achievements, important fundamental questions remain to be answered, especially on the role of phenotypic heterogeneity and how it impacts the response to, as well as how it is changed by, treatments by death ligands [16]. Why do isogenic cells respond differently to the same amount of death ligand? Indeed, it is often the case that not all treated cells die, and when cells die, their death times are very heterogeneous. How different from dying cells are surviving cells before treatment (i.e. *why* cells survive)? How different are surviving cells from what they were just before treatment (i.e. who *are* survivors)? Can these differences explain the decreased efficiency of subsequent treatments (i.e. *what* make cells more resistant)? Importantly, the two last questions, while critical for understanding the efficiency of treatments, are starting to be addressed only since very recently [17, 18]. Here, we review the contributions of system modeling studies to our understanding of receptor-induced apoptosis with a specific focus towards those important questions. We do not aim to exhaustively describe all the modeling work done on receptor-induced apoptosis. Rather, we describe a few key studies that are highly illustrative of how system modeling approaches can provide decisive insights.

## 2. Modeling the biochemistry of receptor-induced apoptosis

## 2.1 Early efforts: from known players and reactions to a system model

Many proteins playing a key role in receptor-induced apoptosis are known since decades, together with a qualitative picture of how they interact, either to convey external death signals to promote the activation of the core executioners of apoptosis or on the contrary to act as 'inhibitors' or 'blockers' of death signaling. Figure 1 provides a schematic overview of apoptotic pathways.

Despite this qualitative knowledge, how precisely cell response emerges from protein interactions in different cell lines and in response to stimulations of different strengths was not well understood. This led Fussenegger and colleagues [19], and later Eissing and colleagues [20], to quantitatively interpret such qualitative schemes and translate them into mathematical models describing the kinetics of the underlying biochemical reactions using the simplest quantitative mathematical framework, ordinary differential equations (ODEs). Assuming specific values for parameters (reaction rate constants and protein initial concentrations) and specific initial conditions (initial protein concentrations), these models can be used to simulate the temporal evolution of molecular species concentrations.

These early studies did not quantitatively compared simulation results to data. Their explanatory power was therefore not well established. Still, by studying how simulated cell behaviors depend on the different parameters, these models provided interesting qualitative insights on the structure of the pathways, that is, on the molecular implementation of receptor-induced apoptosis. For example, Eissing and colleagues rightfully concluded from their model that there must be a caspase-8 inhibitor to allow for both 1) fast kinetics of apoptosis at sufficient stimulation levels and 2) the existence of a threshold stimuli intensity below which apoptosis is not triggered [20].

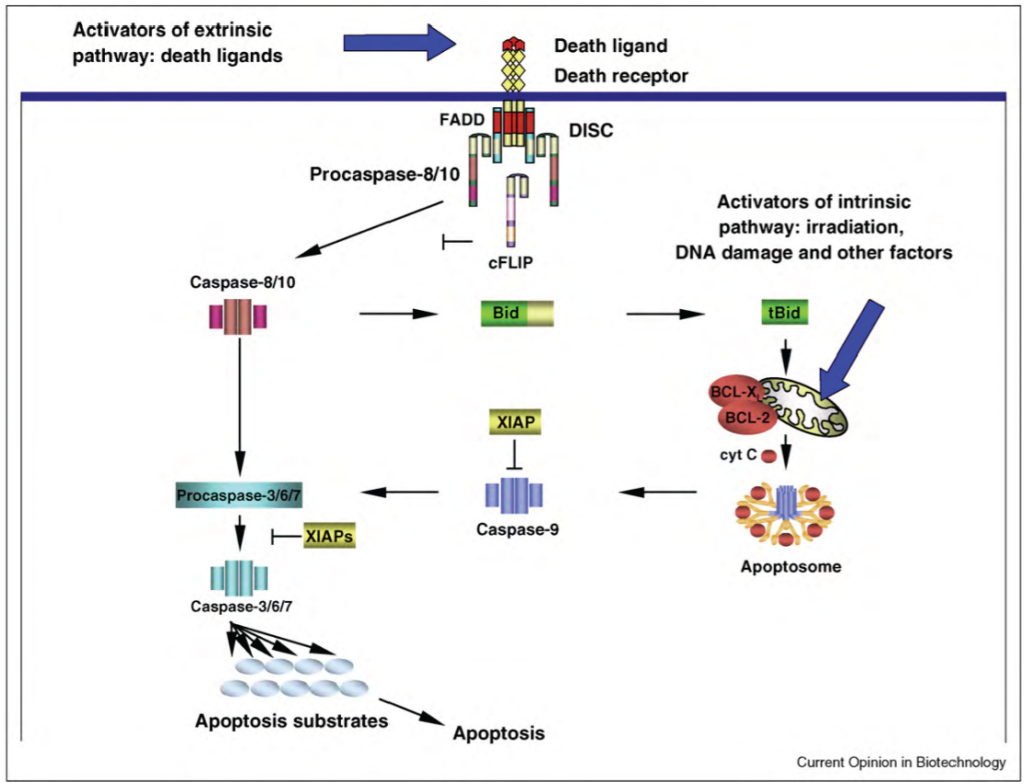

**Figure 1: Schematic representation of extrinsic and intrinsic apoptosis pathways at the molecular level.** Only the main actors and interactions are represented. Death ligands bind their cognate receptors and promote assembly of DISCs complexes that can lead to the activation of the critical initiator caspases caspase-8/10. Initiator caspases can activate effector caspases either directly or by promoting the activation of the mitochondrial apoptosis pathway via Bid. This realizes a connection with the intrinsic apoptosis pathway, also activating the mitochondrial pathway and eventually activating effector caspases. Reprinted from [11], with permission from Elsevier.

## 2.2 Tight integration of kinetic modeling and quantitative experimental data revealed key mechanistic features of receptor-induced apoptosis

After those early efforts, several groups employed approaches that integrated more tightly the construction and analysis of ODE models of receptor-induced apoptosis with experimental data. These approaches have been particularly fruitful. Indeed, they revealed several key mechanistic features of receptor-induced apoptosis.

A pioneering work for the systems biology of apoptosis is the study of CD-95 induced apoptosis by Bentele et al. [21]. The authors constructed an ODE-based kinetic model combining mechanistic and 'black box' reactions. Their initial model contains 41 species, 50 unknown parameters and is notably detailed regarding reactions taking place at the death-inducing signaling complex (DISC): the requirement for the recruitment of two pro-caspase-8 molecules for their activation, as well as the competitive recruitment of cFLIP, were detailed. In order to reduce the risks of over-fitting and detect parameter non-identifiability, the authors used a sensitivity analysis approach to reduce model complexity before testing the model against data. In order to test the model, they obtained quantitative data characterizing the kinetics of caspase activation for several stimulation strengths. The data consisted in quantitative western blots corresponding to protein concentrations averaged over the cell population. The model predicted a threshold for ligand concentration below which no death should be seen, and this prediction was validated experimentally. In the model, the existence of this threshold is caused by cFLIP, which incorporates into newly assembled DISCs and thus blocks pro-caspase-8 processing. Hence, downstream death signaling only occurs for stimulation doses high enough to enable the assembly of a sufficient number of DISCs, capable of overcoming this blockade. By using an inhibitor of protein synthesis (cycloheximide or CHX) and exploiting differences in protein half-lives (cFLIP is short-lived whereas pro-caspase-8 is long-lived), they decreased cFLIP levels while preserving pro-caspase-8 levels and observed the predicted significant decrease of the threshold needed to obtain cell death.

One main limitation of the approach by Bentele and colleagues is the use of population-level measurements for quantifying caspase activation. It was already known that the kinetics of caspase activation was different in different cells. More precisely, single-cell reporters for probing cytochrome c release [22] and effector caspase activation [23] revealed that these events are rapid and relatively invariant in terms of duration and intensity from one cell to another and for different stimulus, whereas their initiation times are highly variable. An important, although often implicit assumption in kinetic models of biochemical pathways is that they represent reactions taking place in individual cells: an enzyme in one cell does not catalyze reactions in another cell. Therefore, in presence of heterogeneity it is not appropriate to reason in terms of population-averaged quantities. Single-cell reporters enabling to measure the abundance or activity of biochemical species with live-cell imaging are therefore appealing tools to test and interrogate on a proper footing kinetic models. And indeed, single-cell reporters in combination with kinetic modeling revealed a number of key mechanistic features of receptor-induced apoptosis.

The first study integrating kinetic modeling with such single-cell data investigated apoptosis induced by staurosporine [24]. Although staurosporine does not induce apoptosis via death receptors, it triggers MOMP (mitochondrial outer membrane permeabilization) and then a rapid, all-or-none effector caspase activation. These molecular events form the downstream part of both extrinsic and intrinsic apoptosis pathways. The authors focused on the events directly following MOMP, using realistic kinetics of cytochrome c and Smac release and apoptosome formation as inputs to their model, which then predicted effector caspase activation kinetics. The amount of XIAP was found to be a key factor in the kinetics of effector caspase activation following MOMP. Interestingly, the model predicted the existence of a small range of XIAP concentrations for which MOMP is followed by a slow and partial effector caspase activation, a prediction that was then confirmed experimentally.

Albeck and colleagues were the first to integrate, in a single model, initiator caspase activation (via TRAIL binding to death receptors), MOMP regulation and effector caspase activation [25]. The model featured 58 species (native protein and protein complexes) and 70 parameters (Figure 2, top left). Instrumental in their work was the development of a single-cell reporter for initiator caspase activity [26], which showed that this activity slowly rises at a variable rate between cells during the pre-MOMP period, and that despite this increasing activity, no significant effector caspase activity is observed; the latter arises suddenly and completely following MOMP (Figure 2, top right). The model revealed that XIAP enables this all-or-none switching behavior not only by competitive binding of caspase 3, but also by promoting its degradation via the proteasome (Figure 2, top center). Another mechanistic insight brought by this model relates to the role of network topology in generating snap-action behavior at the level of MOMP (Figure 2, bottom): Bax multimerization and mitochondrial transport can quantitatively explain the observed behavior despite the presence of Bcl-2, whereas a simple competition model between activated Bax and Bcl2 could not.

Other remarkable works relying on such an integrated approach of quantitative experiments with kinetic modeling include the investigation of the activation of NF-κB signaling in parallel to death signaling in response to CD95-L exposure [27].

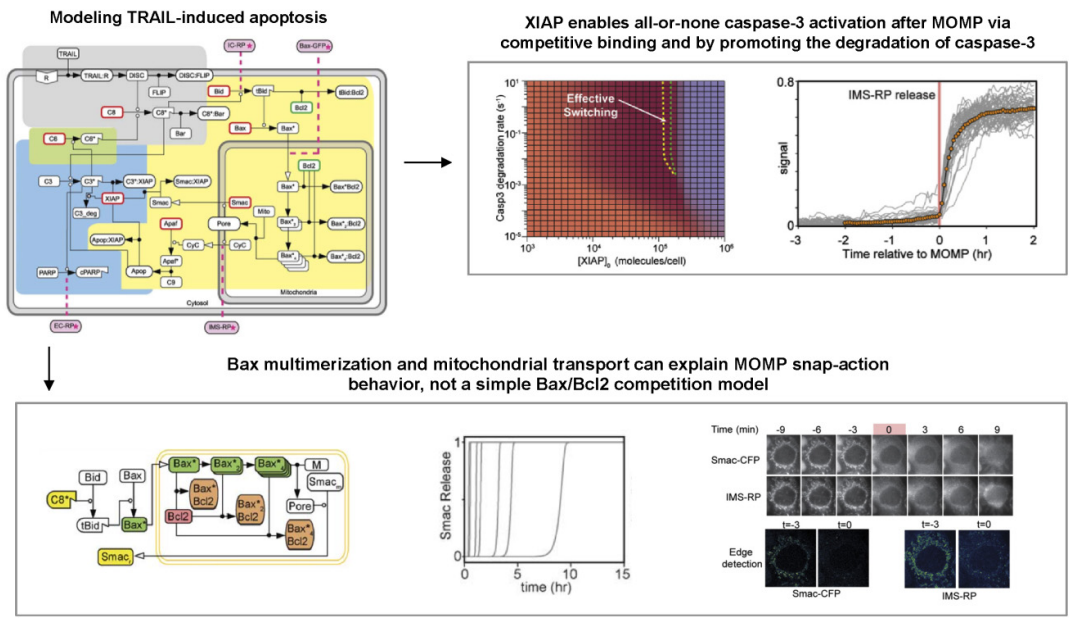

**Figure 2: Key mechanistic features of apoptosis revealed by integrated kinetic modeling and single-cell experiments.** A relatively complete model of TRAIL-induced apoptosis by Albeck and colleagues developed in combination with new single-cell reporters for initiator caspase activity and MOMP allowed new mechanistic insights [25]. For example, XIAP control of caspase-3 activity during the variable pre-MOMP delay does not rely solely on competitive binding but also on its ability to promote caspase-3 proteasomal degradation (right). Another mechanistic insight relates to the role of network topology in generating snap-action behavior at the level of MOMP (bottom). Bax multimerization and mitochondrial transport can explain observed behavior, as opposed to a simple competition model between activated Bax and Bcl2. Figure elements reproduced from [25, 26].

## 3. Modeling populations of individual cells: the role of heterogeneity in protein levels

The previous studies shed light on how snap-action behaviors at the level of MOMP and effector caspase activation enable a tightly-constrained all-or-none control over apoptosis commitment. Such an all-or-none control is probably beneficial at the organism level, because partial effector caspase activation is genotoxic and could result it potential harmful mutations. But why different cells from the same cell line submitted to the same stimulus in the same conditions trigger MOMP after a highly-variable delay from one another? Also, in most studies discussed so far, cells were co-treaded with the protein synthesis inhibitor cycloheximide (CHX). Blocking protein synthesis is interesting to disentangle the influence of signal transduction pathways from the influence of downstream genetic regulations. However, in these conditions, all cells eventually die, whereas in normal conditions, a fraction of the cell population often survives, a property of vital importance in the context of therapy. What are the origins of fractional killing? Are the mechanisms responsible for MOMP timing variability in treatments with CHX also involved in fractional killing without CHX?

An important study from the Sorger group brought key insights into these questions [28]. Using live-cell microscopy, the authors followed the fate of individual HeLa cells after exposure to TRAIL+CHX or TRAIL alone treatments. In both conditions, a significant variability was observed, in death times for TRAIL+CHX treatments, and in cell fate and death times for TRAIL alone treatments. Importantly, to investigate the role of differences in cell state that exist across cells at the time of treatment in determining cell fate and death times, they recorded normal cell proliferation for a duration of about one cell cycle before applying the treatment in order to identify 1) pairs of cells that are sisters and 2) how much time elapsed between their division and treatment. Such lineage information was exquisitely insightful (note that similar experimental observations were made earlier by Rehm et al. [29] and later by Bhola and Simons [30]). First, in the TRAIL+CHX treatment, recently divided sister cells displayed a strong correlation in their death time, despite the high overall variability of death time among cells. This established that 1) death time variability is caused by pre-existing differences, conserved at cell division; and 2) in presence of CHX, TRAIL signaling is almost entirely deterministic (but again, depends on pre-existing differences). In other words, because one could accurately predict the fate of one recently-divided cell by observing the fate of its sister, there is no significant randomness in the signaling reactions taking place between TRAIL+CHX exposure and apoptosis commitment.

This led Spencer and colleagues to the hypothesis that differences in *initial levels* for proteins involved in TRAIL apoptotic signaling are the main determinants of cell fate variability. Mathematical modeling was used to test further this hypothesis (Figure 3). They relied on the same kinetic ODE model (with minor modifications) of the protein-protein reactions mediating TRAIL apoptotic signaling as in their previous study [25]. However, instead of using a single population-averaged value for the initial level of each protein in the pathway, they created an *in silico* cell population by sampling many times protein levels from *distributions,* meant to reproduce the natural variability in protein levels within a population of HeLa cells (some of those distributions were actually measured experimentally using immuno-fluorescence and flow cytometry). Good agreement between model predictions and data for TRAIL+CHX treatments was then obtained (Figure 3), therefore supporting that in these conditions initial variability in protein levels are the main determinants of the observed death time variability.

Additionally, when considering pairs of sister cells born long before TRAIL+CHX treatments, the correlation between their death times continuously decreases, showing that the cell determinants setting this death time fluctuate over time with a timescale of the order of a cell cycle. Notably, protein levels in human cells have been shown to fluctuate with similar time scales [31]. It is therefore probable that the natural slow fluctuations of protein levels are responsible for the decorrelation of sister cell fates after their division. For TRAIL alone treatments, a similar effect is seen, but firstly the correlation for recently-divided cells is reduced compared to TRAIL+CHX treatments, and secondly this correlation decreases markedly faster with sister cells' age. Because the primary effect of CHX is to block protein synthesis, this also strongly suggests that cell fate variability in TRAIL-induced apoptosis originates from synthesis-induced fluctuations in protein levels.

Note that by nature, the model used by Sorger and colleagues cannot account for the sister cell data, a limitation inherent to all deterministic models in which cell-to-cell differences are *static*, that is, cell-to-cell differences are modeled by distributions of values for initial protein concentrations or for time-invariant parameters. Such models do not explain how cell-to-cell variability can be *generated,* which is indispensable if re-establishment of cell-to-cell variability after TRAIL application should be understood. A prime candidate for the (re)generation of cell heterogeneity is stochastic protein fluctuations that are missed out in the previous approach. Note also that no attempt to reproduce cell fate variability data in TRAIL alone treatments, a critical observation, was made. One can cite two reasons. Firstly, the model was trained against TRAIL+CHX data, removing the influence of many parameters constraining protein production. Secondly, fractional killing was thought to result mostly from the activation of survival pathways and these pathways were not included in the model.

Among the other modeling studies investigating cell-to-cell variability in receptor-induced apoptosis, it is worth mentioning the work by Toivonen and colleagues [32]. These authors extended the model of CD-95L induced apoptosis from [21] with variable synthesis and degradation rates for the short-lived protein c-FLIP, and their analysis predicted c-FLIP targeted degradation as being a fundamental determinant of death receptor responses, in agreement with experimental observations.

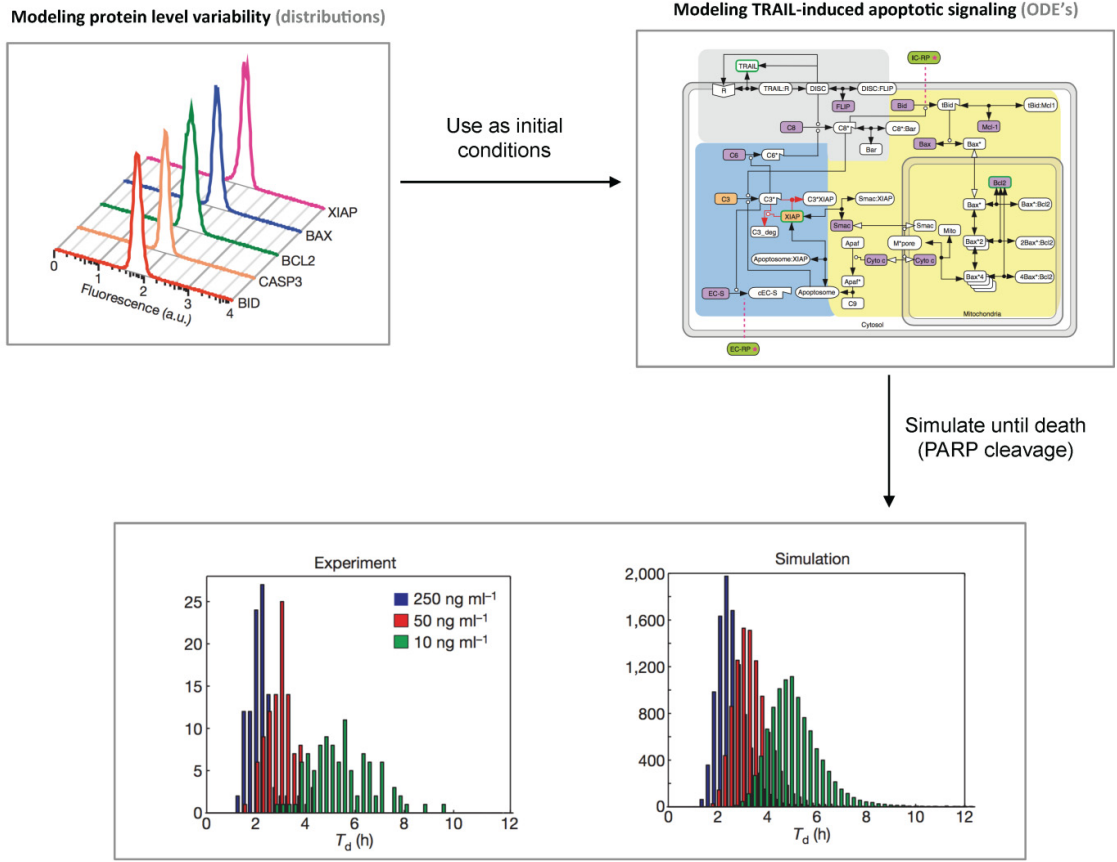

**Figure 3: Initial variability in protein levels explains variability in the timing of death.** Spencer and colleagues combined a previously-proposed ODE model of TRAIL-induced apoptosis signaling with the use of distributions for the initial values of proteins levels as could be measured by immunochemistry and flow cytometry to represent the heterogeneity in protein levels existing in cell populations [28]. Then they recorded the variability in the timing of death for different doses of TRAIL and in presence or absence of cycloheximide within the *in silico* cell population. The resulting distributions closely resemble the distributions obtained from experimental data. Figure elements reproduced from [28].

## 4. Modeling different cell lines and their different sensitivities to receptor-induced apoptosis

From the discussion in the previous section, we have seen that differences in protein levels could be a main determinant of cell fate differences. Thus, knowing the expression levels of the proteins involved in extrinsic apoptosis in a given cell line could help predicting its sensitivity to different death ligand stimulations. Stated differently, we adopt here the viewpoint in which cell lines do not differ by the *topology* of their pathways but rather by the *levels* or more specifically by the *distributions* of the proteins involved in these pathways. This idea motivated another study by the Sorger group [33]. Using their previous model as a reference for the rates of biochemical reactions of extrinsic apoptosis, they studied the biochemical basis of the Type I / Type II behaviors. Type I (II) behavior refers to cells that do not require (do require) MOMP to commit to apoptosis after death ligand stimulation. As a consequence, a strong over-expression of Bcl2 proteins renders cells resistant to death ligand stimulation only in type II cells.

The authors could successfully classify the Type I/II behavior of a set of cell lines solely based on the expression levels of the proteins involved in apoptotic signal transduction. Their approach was based on direct finite-time Lyapunov exponent (DLE) analysis, which measures the influence of changes in initial protein concentration on the future states of the system. More precisely, when computing DLEs for different initial conditions, they obtained a narrow region of high DLE values, i.e. a region where small changes in initial conditions lead to large deviations in cell state after stimulation, separating two large regions having comparatively low DLE values (Figure 4, bottom left). When positioning cell lines on this space based on measured expression levels, they found that Type I and Type II cell lines were on opposite sides of the high-DLE region, while cell lines exhibiting mixed behaviors were close to it.

One limitation of the DLE analysis is that the DLE is a number that is difficult to interpret. It reflects a sensitivity of the future states to the initial conditions, but it does not give information about what is perturbed in the states. In addition, one has to choose a time horizon to compute DLEs, which might have a strong influence on the results. We have therefore proposed another approach, based on Signal Temporal

Logic (STL) instead of DLE [34]. Temporal logics are flexible property specification languages that allow describing expected features of behaviors. Experimentally-observed behaviors are explicitly encoded in STL. This approach allowed us to discover that the notion of Type I and Type II has limits, as there exist several interpretations of being a Type I or a Type II cell which are not equivalent (Figure 4, bottom right).

The idea that differences in protein expression levels between cell lines could predict differences in response to death ligand stimulation from a mechanistic model of extrinsic apoptosis was also used in other studies [35]. Recently, a similar approach was applied to patient-derived cell lines to predict their sensitivity to treatment [36], although here they used statistical modeling (rather than mechanistic modeling) to map expression profiles to sensitivity.

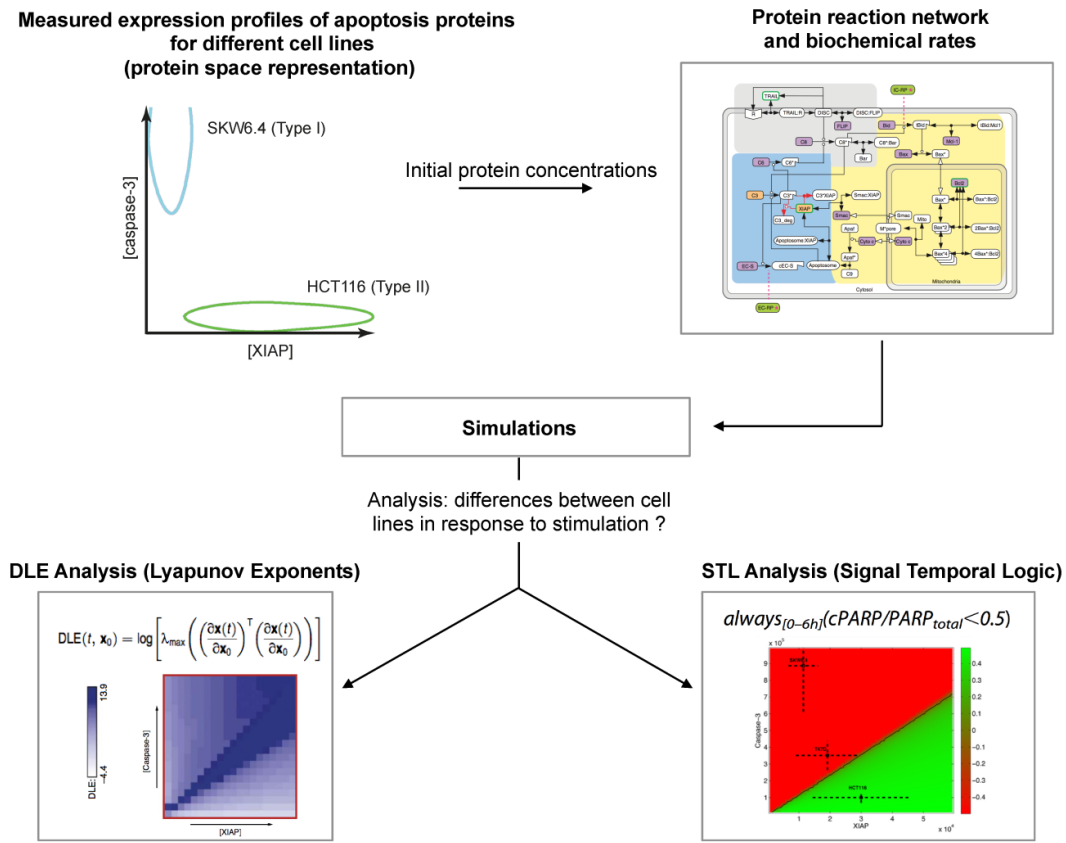

**Figure 4: Simulating the differential sensitivity of cancer cell lines to receptor-induced apoptosis.** Different cell lines express extrinsic apoptosis proteins at different levels (leading to different distributions of single-cell level expression), and those differences a priori impact on their response to receptor stimulation. Aldridge and colleagues [33] used Lyapunov exponent analysis to study and demonstrated its ability to classify and compare different cell lines. Stoma and colleagues [34] proposed an alternative to Lyapunov exponents, Signal Temporal Logic (STL), that allows to formally encode behavioral differences as measured by various experimental assays. Figure elements reproduced from [33, 34].

## 5. Modeling fluctuations of protein levels to extend the temporal scope of existing models

Until now, the modeling approaches we have discussed represent the naturally-arising differences in protein levels between individual cells of a given cell line by static distributions, and such distributions are then used as initial conditions for deterministic models of extrinsic apoptosis signaling (see Section 3). In particular, one of the most important mechanisms that generate these distributions, the burstiness of gene expression and therefore the stochastic nature of protein turnover, is not accounted for in the above-mentioned models. Whether such fluctuations are responsible for the observed decrease of the correlation of death times of sister cells with their age at treatment as discussed in Section 3 is an interesting question. Not accounting for protein fluctuations fundamentally limits the temporal scope of a given protein-protein interaction model, even if its kinetic parameters are appropriately constrained. With this approach, we addressed the question of what *are* surviving cells after treatment, and hence what will be their resistance to future treatment applications (Figure 5, top) [18]. It relies on modeling stochastic gene expression (stochastic switches of the promoter between an active and an inactive transcriptional state, and stochastic production and degradation of the mRNA) and protein turnover for *all* (native) proteins appearing in the model. As a result, protein levels slowly fluctuate in each individual cell such that, overall, the distributions of the protein concentrations in the whole cell population are the ones observed in [28]. This means that the naturally-occurring cell-to-cell variability, previously accounted by pre-determined distributions for initial protein concentrations, is now an emerging property of the model.

While such model extension a priori introduces many unknown parameters, we found that using simple constraints from the literature on parameter values one readily obtains good approximations of protein fluctuations for most proteins. Only short-lived proteins necessitate particular attention. This finding is a cornerstone of the approach, as it allows to strongly reduce the number of unconstrained parameters, facilitating exploration of the parameter space and reducing the risks of over-fitting. We applied this modeling approach to extend the model of TRAIL-induced apoptosis used by Spencer and colleagues. Among the 17 native proteins appearing in the model, only gene expression parameters for Flip and Mcl1 (known to be very short-lived at both the mRNA and protein levels) were used for fitting the data, while standard parameter constraints were used for all others. The model could quantitatively fit cell death distributions and cell survival fractions for both TRAIL+CHX and TRAIL alone treatments (Figure 5, middle, left). Moreover, sister cell data (decrease in their death time correlation as they age) that have not been used to fit our model could be predicted (Figure 5, middle, right), thus validating the approach.

The finding that cell survival does not require TRAIL-induced activation of survival pathways, but can occur solely from the interplay of stochastic gene expression, fast turnover of certain anti-apoptotic proteins, Flip and Mcl1, and rapid degradation of activated forms challenges the classical view about the role of survival pathways in response to TRAIL [37]. While it does not mean that survival pathways do not play a role in cell survival after TRAIL exposure, our results strongly suggest that they are not the sole contributors to cell survival.

Finally, because the model can predict changes in cell states (ie cell protein content) of the population caused by a first treatment as well as the recovery of cells to their normal states (ie initial protein distributions) after treatment based on stochastic gene expression and protein turnover, the efficiency of a second treatment as a function of the time between treatments could be investigated. In agreement with data, simulations showed a marked but transient increase of the population resistance after a treatment (Figure 5, bottom). We therefore provide a simple mechanistic explanation to the observed reversible resistance of cells to repeated treatments.

**A** Approach: accounting for stochastic gene expression in signal transduction models

- Systematic: fluctuations of all proteins are modeled
- Parsimonious: standard parameter constraints used for long-lived proteins, specific attention only given to short-lived protein

Example: portion of the extrinsic apoptosis pathway

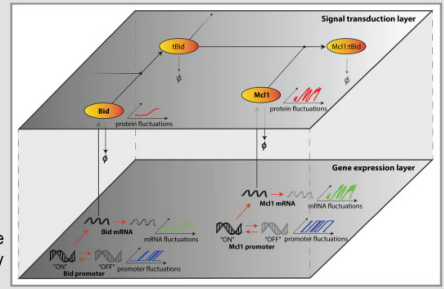

**B** Model explains cell fate variability and predicts transient cell fate inheritance

Description of the experiments by Spencer et al., 2009

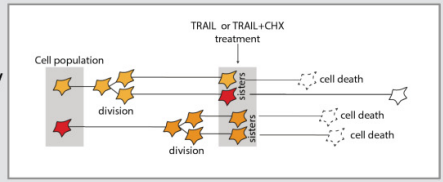

Fitting on cell fate variability data

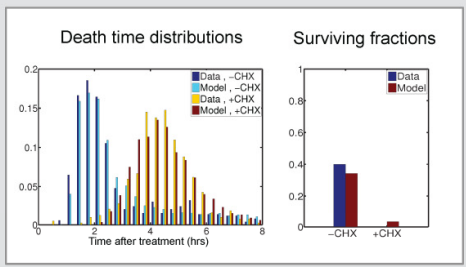

Validation using sister cell data →

Death time correlation between sister cells

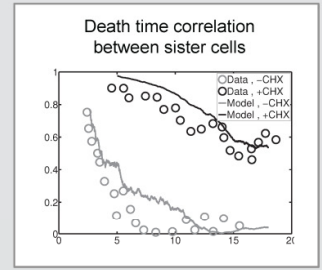

**C** Model predicts transient resistance acquisition after treatment

In-silico experiment: repeated TRAIL treatment

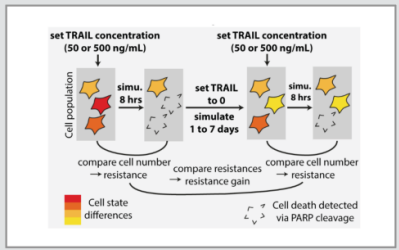

Model prediction

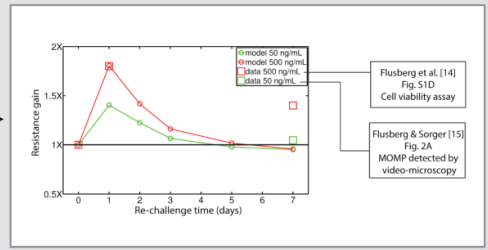

**Figure 5: Systematic, parsimonious modeling of stochastic gene expression together with TRAIL apoptotic signaling explains fractional killing and predicts transient cell fate inheritance and transient resistance acquisition.** (A) Schematic description of the modeling approach. (B) Results of the approach when applied to model and data of Spencer and colleagues [28]. (C) Simulation of consecutive TRAIL treatments reproduces the observed transient resistance acquisition [18]. Figure elements reproduced from [18].

## 6. Conclusions and perspectives

In this chapter, we have surveyed how system modeling of receptor-induced apoptosis has been instrumental in improving our understanding of this process at several levels: the molecular level, the level of cellular decisions between life and death, and the level of cell populations exhibiting various degree of resistance as a function of their protein expression profiles or their treatment history.

More precisely, ordinary differential equations models recapitulating known reactions between proteins during apoptosis signaling are useful when compared to short-term (a few hours) population data about protein level and state kinetics (Section 2). They allow verifying that the structure of known reactions is compatible with what is experimentally observed and can provide estimates of the associated biochemical rates (although parameter non identifiability often prevents the obtention of unique estimates). Comparing model predictions with population data has fundamental limitations, and comparison to single-cell data (obtained by means of cell-level reporters for well-defined biochemical activities or events) is a priori more meaningful. Indeed, in the context of receptor-induced apoptosis, it has revealed important kinetic features regarding MOMP regulation and effector caspase activation (Section 2).

Still, there are two difficulties arising when comparing ODE models of protein-protein reactions with such single-cell data. First, protein synthesis, which is noisy and hence generates differences from cell to cell, can have an impact on signaling dynamics at the protein level. This effect can only be temporally mitigated by using protein synthesis inhibitors like cycloheximide. Second, model predictions depend on initial conditions, such as the pre-stimulation levels of the protein involved in receptor-induced signaling. We have seen that most of the variability in death timing following TRAIL (and cycloheximide) treatment can be explained when realistic random distributions of initial protein levels are used as initial conditions for an ODE model of TRAIL-induced apoptosis signaling (Section 3). This result is important as it demonstrates that TRAIL-induced apoptosis signaling is not intrinsically noisy, and that cell state (i.e., the levels of apoptosis proteins) differences at treatment time are a major determinant of cell fate variability. Indeed, we have seen that the protein expression profiles of different cell lines can inform about their sensitivity to extrinsic death stimulation when used as initial conditions of a single (i.e. the same for all cell lines) ODE model of apoptotic signaling (Section 4).

However, while cell fate is almost fully determined by cell state at treatment time when protein synthesis is blocked, it is only partially the case in normal treatment conditions, in which survival of a fraction of the population is often observed. Indeed, protein synthesis can interact with receptor-induced signaling and steer cell fate in one direction or another. In section 5, we have seen how systematic but parameter-parsimonious modeling of stochastic gene expression within ODE models of signal transduction dynamics can explain important observations on the dynamics of cell-to-cell variability in TRAIL-induced apoptosis. This approach allows extending the temporal scope of ODE models of receptor-induced apoptosis. This is required to investigate the response of cell populations to multiple treatments separated in time, for which resistance acquisition is very often observed. An important prediction of such models is that transient resistance acquisition can occurs in absence of stimulus-induced pro-survival transcriptional activity.

Despite those promising advances, many questions remain without clear answers. For example, while the important role in cell survival of the targeted degradation of many pro- and anti- apoptotic proteins is increasingly recognized [38], accurate estimates of the corresponding rates are not available, and to which extent those rates fluctuate in single cells and vary from cell to cell is not known. Experiments using proteasome inhibitors are difficult to interpret because they have a global (but not necessarily identical) effect on all degradation rates. More targeted approaches (for example using specific single-cell reporters) could be very useful to better understand the role of targeted degradation in receptor-induced apoptosis.

While current models of receptor-induced apoptosis can be quite large (up to 100 species and reactions), they are often omitting structural details either because those details are not understood very well or because a simplifying representation is deliberately preferred. For example, the ligand-induced receptor clustering at cell surface, the processing of caspase-8 at the DISC, the role of the different Flip isoforms in that processing, and the interactions of all MOMP regulators at the mitochondrial surface, are generally significantly simplified. Such simplified representations can be accurate and therefore sufficient to address many questions. Still, to test and improve our molecular-level understanding of receptor-induced apoptosis, more detailed mechanisms can be introduced into existing models, and model predictions can be compared to new data generated with adequate tools (such as relevant single-cell reporters). Without the 'right' data, increasing model complexity is probably vain.

Finally, while mathematical models of receptor-induced apoptosis start to address the question of long-term behavior of cell populations repeatedly treated by death receptor agonists, the amount and quality of corresponding experimental data is very scarce. Most in-vitro studies still only measure the efficiency of one-time treatment to model cell lines, and repeated treatments are only seen in mouse xenografts studies, in which time points and measurements are limited, and many effects related to the in-vivo context can affect the response. Quantitative population dynamics data (i.e., cell proliferation and death rate as a function of time) for cell lines cultured in-vitro and submitted to repeated treatments could prove very useful to better understand resistance acquisition, and to map it to molecular mechanisms with the aid of mathematical models. Also, the potential impact of spatial organization of the cells in tumors may generate further inhomogeneities that are not captured by models disregarding space, hence a full understanding will eventually have to explore the possible effects of space.

## Acknowledgments

This work was supported by the research grants Syne2arti ANR-10-COSINUS-007 and Investissements d'Avenir Iceberg ANR-IABI-3096 from the French National Research Agency.